\documentstyle[aps,preprint,epsf,prb]{revtex}
\tightenlines
\begin{document}

\title{Correlation studies of open and closed states
fluctuations in an ion channel: Analysis of ion current through a
large conductance locust potassium channel}

\author{Zuzanna Siwy$^{1,2}$\footnote{Email address:
siwy@zeus.polsl.gliwice.pl}, Marcel Ausloos$^3$\footnote{Email address:
marcel.ausloos@ulg.ac.be}, Kristinka Ivanova$^4$\footnote{Email address:
kristy@essc.psu.edu}}

\address{$^{1}$Department of Physical Chemistry and Technology of Polymers,
Silesian University of Technology, 44-100 Gliwice, Poland\\
$^{2}$Materials Research Department, Gesellschaft fuer
Schwerionenforschung, 64291 Darmstadt, Germany\\
$^{3}$SUPRAS \& GRASP, Institute of Physics B5, University
of Li\`ege, B-4000 Li\`ege, Euroland\\
$^{4}$ Pennsylvania State University,
University Park, PA 16802, USA}

\maketitle

\begin{abstract} 

Ion current fluctuations occurring within open and closed states
of large conductance locust potassium channel (BK channel) were
investigated for the existence of correlation. Both time series,
extracted from the ion current signal, were studied by the
autocorrelation function (AFA) and the detrended fluctuation
analysis (DFA) methods. The persistent character of the short- and
middle-range correlations of time series is shown by the slow
decay of the autocorrelation function. The DFA exponent $\alpha$
is significantly larger than 0.5. The existence of
strongly-persistent long-range correlations was detected only for
closed-states fluctuations, with $\alpha=0.98\pm0.02$. The
long-range correlation of the BK channel action is therefore
determined by the character of closed states. The main outcome of
this study is that the memory effect is present not only between
successive conducting states of the channel but also independently
within the open and closed states themselves. As the ion current
fluctuations give information about the dynamics of the channel
protein, our results point to the correlated character of the
protein movement regardless whether the channel is in its open or
closed state.
\end{abstract}

%\vskip 1.0cm 



\section{Introduction}

Ion transport through biological channels is a phenomenon which
has attracted attention of biophysicists and biologists for many
years \cite{hille}. The main reason of this profound interest is
that ion channels, abundant in most, if not all, eukaryotic cells,
are involved in many physiological processes such as secretion,
regulation and membrane potential, signal transduction and
osmoregulation \cite{hille,schrempf}. A breakthrough in such
studies was brought about by the patch clamp technique, which
enables to measure single-channel ion currents with millisecond
time resolution and therefore probes the dynamics of the channel
protein \cite{neher}. The tiny, picoamp current through an
individual ion channel "clamped" at constant voltage does not lead
to a constant signal but presents a peak-valley landscape pattern
of high and low current values. The channel's opening state can be
determined on the basis of the ion current: a low current
corresponds to a closed channel state, while high current values
indicate an open state. The analysis of ion current recordings has
been found to be a tremendous challenge for many theoreticians.
Usually the data were analysed in terms of models and tools based
on the assumption that the channel kinetics is a Markov process
\cite{gardiner} over a small number of discrete states
\cite{hille,neher,mozrz}. There is, however, clear evidence that
the behavior of ion current is non-Markovian \cite{petra,timmer}.
For example, the method of testing Markovianity, based on the
Smoluchowski-Chapman-Kolmogorov equation \cite{ful} distinctly
revealed a non-Markovian action of the large conductance locust
potassium channel (BK channel). Dwell-time series of the same
system subjected to the (i) autocorrelation, (ii) Hurst and (iii)
detrended fluctuation (DFA) analysis showed the existence of
long-range memory effects in the BK channel \cite{siwy}.

Previously, the focus of attention has been on the character of the
transitions between open and closed states of the channel without
detailed studies of fluctuations occurring within one state. If
large conformational changes of the channel protein result in
channel closing and opening, observed as low and high values of the
current, respectively, then the fluctuations of closed and open
states can reflect the protein conformational substates. The
existence of a spectrum of protein substates occurring on a wide
time scale range is supported by strong experimental evidence, and
has been widely discussed
\cite{karplus,lieb1,lieb2,lieb3,salman1,salman2,braun}. It would
therefore be interesting to answer the question whether this
channel action is also governed by a deterministic force, seen as a
correlation between the substates.

A new way to examine this problem appeared after a thorough study
of the effect of trends and noise in the detrended fluctuation
analysis (DFA), as presented by Hu et al. \cite{hu}. These authors
showed an elegant and rather thorough analysis of time series
containing signals with different correlation characteristics. As
ion current recordings can be also treated as a superposition of
two signals, i.e. closed and open states dynamics, we have applied
this DFA technique to a time series of ion current through BK
channel. The experimental recording has been divided into two
signals, containing closed and open state fluctuations,
respectively. We consider therefore the closed (open) state as a
whole with the opposite state as an "interruption". It enabled us
to check the existence of long time-lag memory effects occurring in
spite of the gaps in channel activity. The resulting time series
have been subjected to both a DFA and an autocorrelation function
analysis (AFA).

\section{Experimental design}

\begin{figure}[ht] \begin{center} \leavevmode \epsfysize=8cm
\epsffile{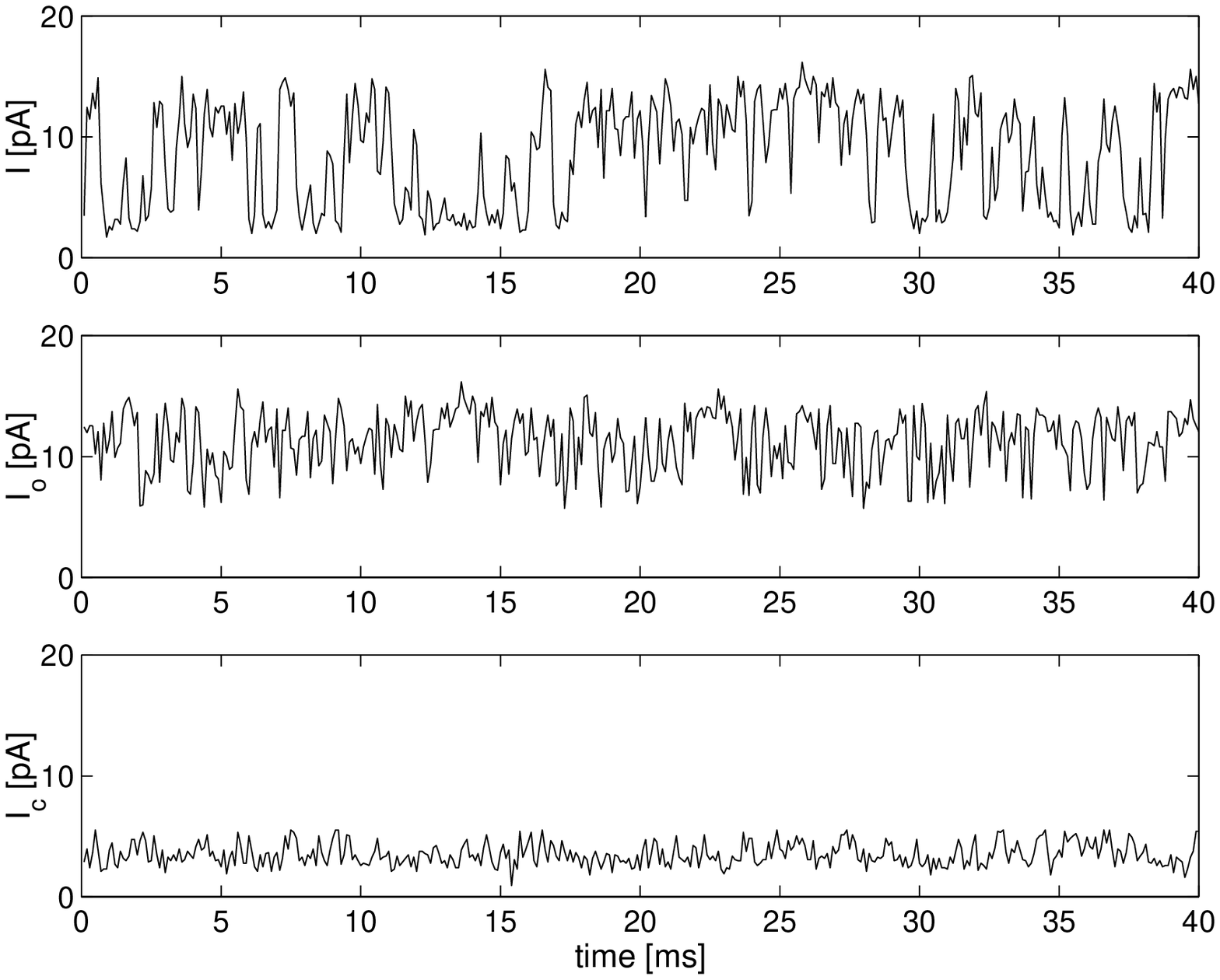} 
\caption{Ion current time series through a large conductance
locust potassium channel, recorded at the pipette potential 60 mV
(top), together with extracted series of open $I_{o}$ (middle),
and closed $I_{c}$ (bottom) states fluctuations. The closed and
open state fluctuations were found on the basis of the threshold
current $I^*=5.6\pm0.2$ pA, separating two modes in the ion
current probability density function (PDF) \cite{mws1}. A given
ion current point is categorized as belonging to an open state if
$I>I^*$, otherwise it belongs to the closed state time series.}
\label{fig1}
\end{center}
\end{figure}

Data sets from cell-attached patches of adult locust (Schistocerca
gregaria) extensor tibiae fibers \cite{ful,gorcz} are discussed.
The potassium current through a voltage-sensitive large
conductance locust potassium channel (BK channel) was measured by
the patch clamp technique \cite{neher} with a sampling frequency
$f_{ex}=10$ kHz and at a 60 mV voltage. Details of the
experimental set-up and recordings of the data have been given in
\cite{ful,mws1}. The analyzed ion current data consist of one
record of 25 s duration, composed of 250~000 points (see
Fig.~\ref{fig1}). All points in the time series have been first
divided into two groups, to be called closed and open, depending
on their values in relation with the threshold current
$I^*=5.6\pm0.2$ pA, separating two modes in the ion current
probability density function (PDF). The value of $I^*$ has been
determined by (i) approximation of PDF by means of kernel density
estimator technique, and (ii) presenting the resulted function in
double-logarithmic coordinates \cite{mws1}. The PDF has been
treated as a superposition of two unimodal densities with distinct
power laws, whose intersection determined the threshold $I^*$
separating closed and open states. A given ion current point is
categorized as belonging to an open state if $I>I^*$, otherwise it
belongs to the closed state. As we want to investigate correlation
properties of closed and open states separately, the fluctuations
of the two states have to be extracted from the ion current time
signal. This was achieved using the procedure discussed in
\cite{hu}, based on cutting out "bad", i.e. non-interesting parts
of a time series: for extracting the closed state signal, all
points belonging to the open states were cut out and the resulting
segments 'stitched' together. The same procedure has been applied
for getting open states fluctuations: after cutting out closed
states, the open states points were joined together. The resulting
time series of closed $I_{c}$ and open $I_{o}$ states fluctuations
contained 128 760 and 121 240 points, respectively; the signals
are presented in Fig.~\ref{fig1}.

\section{Results and discussion}

\subsection{Detrended fluctuation analysis} The detrended fluctuation
analysis
is a tool for investigating correlations in a self-similar time
series with stationary increments \cite{hu,ausloos1,ausloos2}. It
provides a simple quantitative parameter - the scaling parameter
$\alpha$, which is a signature of the correlation properties of the
signal. The advantages of DFA over many methods are that it permits
the detection of long-range correlations embedded in seemingly
$non-stationary$ time series, and also that inherent $trends$ are
avoided at all time scales. The DFA technique, similar to the Hurst
analysis, can be applied to a random walk "mimicking" the
cumulative time series. The DFA technique consists of dividing a
time series $x(t)$ of length $N$ into $N/n$ nonoverlapping boxes
(called also windows), each containing $n$ points \cite{stan5}. The
local trend $z(t)$ in each box is defined to be the ordinate of a
linear least-square fit of the data points in that box. The
detrended fluctuation function $F_k(n)$ is then calculated
following: \[
F^2_k(n)=\frac{1}{n}\sum_{t=kn+1}^{(k+1)n}{\left(x(t)-z(t)\right)^2},
\hspace{10mm} k=0,1,2,...,\left(\frac{N}{n}-1\right). \] Averaging
$F^2_k(n)$
over the $N/n$ intervals gives the mean-square fluctuations
$f(n)=\sqrt{\left<F^2(n)\right>}$ as a function of window size $n$. The DFA
exponent $\alpha$ is obtained from the power law scaling of the function
$f(n)$
with $n$, and represents the correlation properties of the signal:
$\alpha=0.5$
indicates that the changes in the values of a time series are random and,
therefore, uncorrelated with each other. If $\alpha<0.5$ the signal is
anti-persistent (anti-correlated), while $\alpha>0.5$ points to the positive
persistency (correlation) in the signal.

\begin{figure}[ht] \begin{center} \leavevmode \epsfysize=8cm
\epsffile{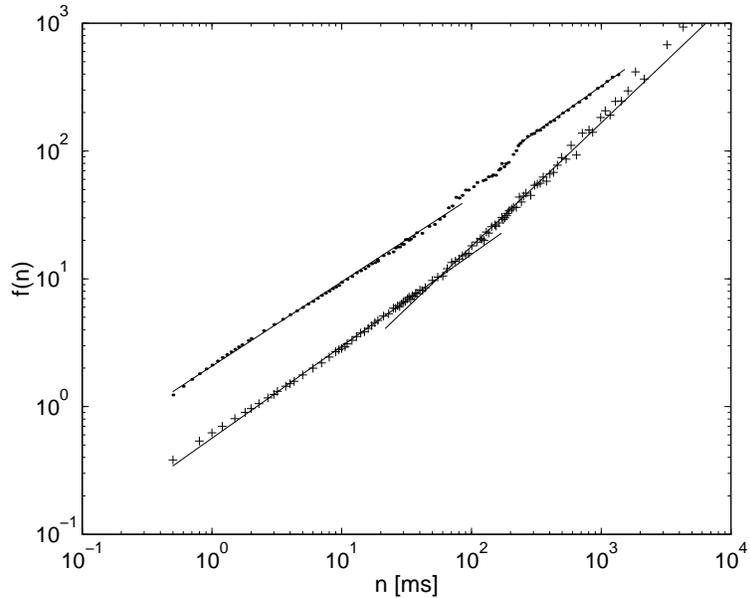} \caption{The DFA analysis for closed (crosses) and open (points) states
fluctuations (see Fig. 1); the mean-square fluctuations function $f(n)$ is
plotted as a function of the window size $n$.}
\label{fig3}\end{center}
\end{figure}

\begin{figure}[htb]
\begin{center}
\leavevmode
\epsfysize=8cm
\epsffile{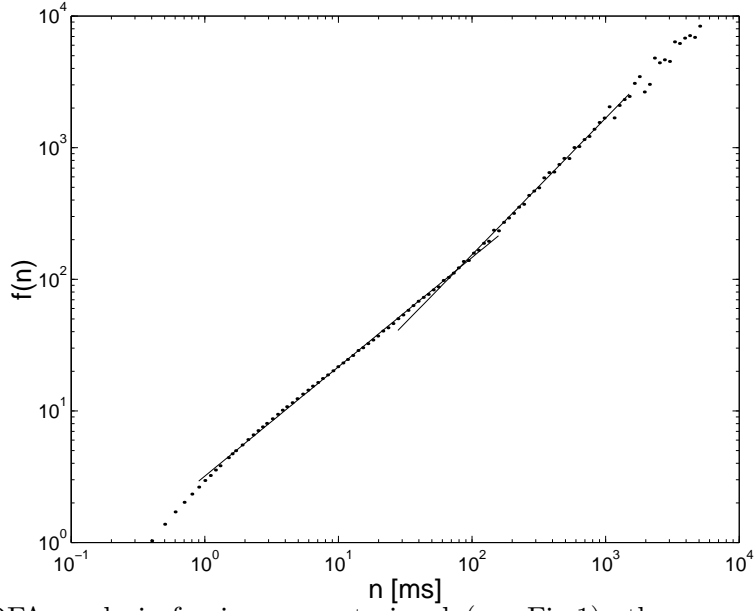}
\caption{The DFA analysis for ion current signal (see
Fig.1); the mean-square fluctuations function $f(n)$ is plotted as a
function of the window size $n$. Linear regression in loglog
coordinates revealed two scaling regions: for $n>70$ ms
$\alpha=0.83\pm0.01$, while for $n>70$ ms exponent $\alpha$ equals
$1.04\pm0.04$.} 
\label{fig2}
\end{center}
\end{figure}

The results of the DFA analysis of cumulative ion current time
series (studied before in \cite{mercik2}) and its two components:
the series of closed and open states fluctuations, are presented
in Figs.~\ref{fig2} and \ref{fig3}. The DFA analysis of the ion
current signal is revisited here. In \cite{mercik2} the average
$\alpha$ for the whole examined time range has been found equal to
$\alpha=0.89\pm0.07$. Here we show that the two significant
regions of DFA scaling can be distinguished, with the threshold at
70 ms. For windows smaller than 70 ms $\alpha=0.83\pm0.01$ while
for $n>70$ ms $\alpha=1.04\pm0.04$. The significance (with 5\%
significance level) of the scaling difference in the two regions
($0.5<n<70$ ms and $70<n<2000$ ms), has been assessed by a t-test.
The paired t-test has been used to show a significant difference
between the two series obtained after fitting the DFA data with
$\alpha=0.89\pm0.07$, and newly found $\alpha$ values,
respectively. As the existence of a threshold may depend on the
order of DFA applied \cite{hu,kantel}, we confirmed that our
results are the same also for DFA of higher orders \cite{hu}.

A power law scaling of the $f(n)$ function for the three signals
has been found over a wide time scale range (three orders of
magnitude); the DFA exponent significantly higher than 0.5, points
to their persistent character. Note that the correlation ranges
within the closed and open states are very similar. The analysis
indicates that the character of channel dynamics within certain
time scale range up to ca. 50 ms is the same, regardless of the
channel conducting state.

The correlation properties of closed states shows crossover of the function
$f(n)$ at app. 50 ms. The function $f(n)$ scales as following:

\[ f_{c}(n)\propto\left\{ \begin{array}{ccl} n^{0.69\pm 0.02} & {\rm
for } & n<50
{\rm ms}\\ n^{0.98\pm 0.02} & {\rm for } & n>50 {\rm ms}\\
\end{array} \right. \]

As the time series contains two closed states of duration longer
than 200 ms, which corresponds to over 2000 experimental points,
we calculated the local $\alpha$ for those closed states. Similar
analysis was not possible for open states as the longest duration
of open state is only $27.2\pm 0.10$ ms, which does not assure a
reliable averaging required in the DFA analysis. For two closed
states of duration 300.8 ms and 202.9 ms, in the range $0.5<n<20$
ms, $\alpha=0.71\pm 0.10$, which is in a very good agreement with
the value determined on the basis of the overall closed state
signal for $n<50$ ms.

The mean-square fluctuation function $f(n)$ for open states reads:

\[ f_{o} (n)\propto\left\{ \begin{array}{ccl} n^{0.65\pm 0.01} & {\rm for }
&
n<80 {\rm ms}\\ n^{0.71\pm 0.01} & {\rm for } & 240<n<1500 {\rm ms}\\
\end{array}
\right. \]

For the range of window size between 80 and 240 ms it was not possible to
determine reliably the scaling exponent for open state fluctuations.

Note, that the crossover of DFA of closed states fluctuations (50
ms) is similar to the crossover found in the correlation between
successive ionic current values (Fig.~\ref{fig2}). For both
signals the threshold determines the onset of strongly persistent
long-range correlations with $\alpha$ close to 1. Interestingly, a
similar scaling behaviour was found in both, the time series of
closed states' durations as well as series consisting of all dwell-times,
closed and open, studied in our previous paper \cite{siwy}. The
correlations between successive durations of closed states are
characterised for $n>70$ ms also by a high $\alpha$ exponent equal
to $0.86\pm0.07$. Similar scaling was found for the series of
dwell-times durations with the threshold at $n=100$. The memory
within and between the closed states as well as between neighbouring
conducting states, might therefore originate
from a similar force(s), one of which is certainly the applied
voltage. The influence of the external field is especially visible
at large time scale ($n>70$ ms): $\alpha$ close to 1 for ion
current signal, closed states, and dwell-times series,
indicates a strongly persistent ion movement through the pore.

We would like to point out that our results, showing persistent
self-similar changes of ion current within closed and open states
of the channel, support fractal models of the ion channel action,
developed by Liebovitch et al. \cite{lieb5,lieb4}. The power law
(or scaling) nature of the ion channel $f(n)$ function implies that
the timing of channel switchings is self-similar therefore
correlated within a certain time scale. It has further consequences
for the state of the channel openings and closings, therefore
rearrangements of the channel protein structure itself. If the
channel protein operates in a fractal mode then the open-closed
rearrangements do not happen all by themselves but rather result
from smaller changes, introduced by the motion of small portions of
the protein. The little pieces of the protein function, according
to such a fractal model, "in a concerted whole" when the channel
switches \cite{lieb5}. Thus, the above results on the persistent
character of the closed and open states fluctuations likely reflect
the persistent, coherent conformational changes of the channel
protein.

\subsection{Autocorrelation function}

\begin{figure}[ht] \begin{center} \leavevmode \epsfysize=8cm
\epsffile{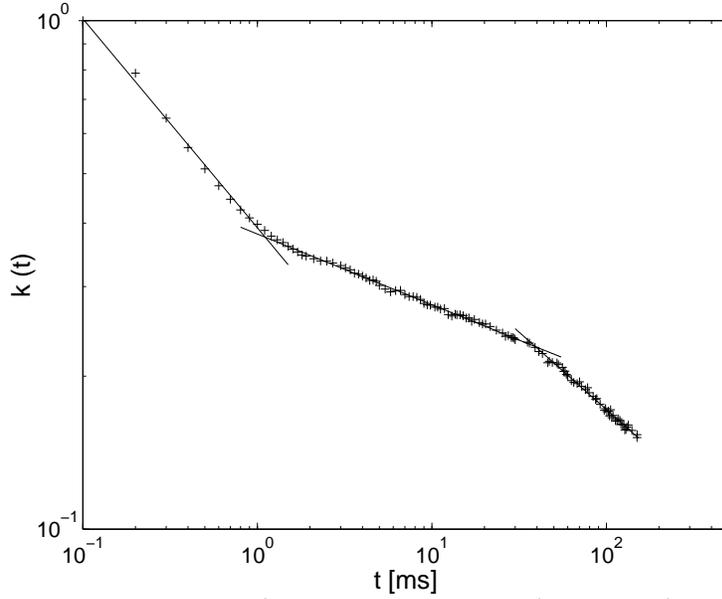} \caption{Double logarithmic plot of the autocorrelation
function of the ion current signal (see Fig. 1) \cite{mws1}). The
power-laws are
$t^{-0.35\pm0.05}$ for $t<1$ ms, $t^{-0.14\pm0.02}$ for $1{\rm\
ms}<t<40$ ms, and
$t^{-0.28\pm0.20}$ for $t>40$ ms \cite{mws1,mercik2}.} \label{fig4}
\end{center}\end{figure}

\begin{figure}[ht] \begin{center} \leavevmode \epsfysize=8cm
\epsffile{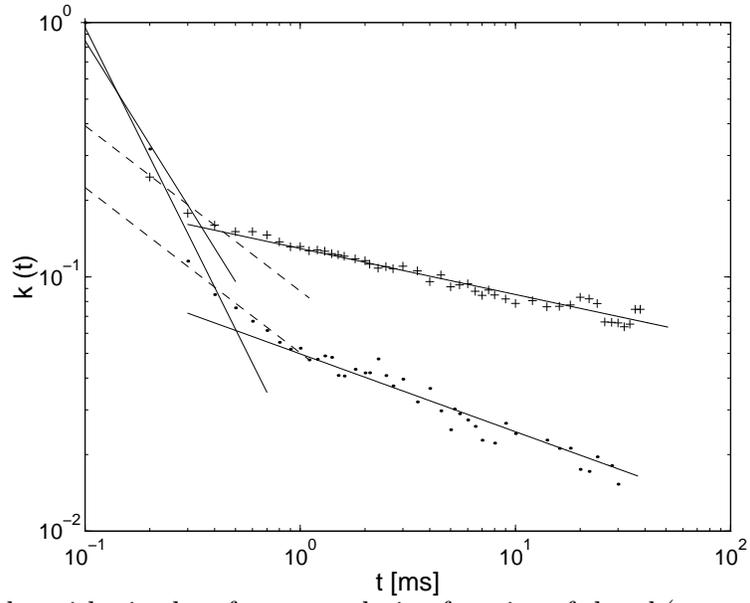} \caption{Double logarithmic plot of
autocorrelation function of closed (crosses) and open (points)
states fluctuations (see Fig. 1). The dashed lines show the
relation $t^{-0.65}$, which indicates the average expected value
of the AFA exponents $\gamma$, calculated from the theoretical
relationship between $\gamma$ and $\alpha$ exponents \cite{bass}.}
\label{fig5}\end{center}
\end{figure}

The autocorrelation function is next used in oder to investigate
the correlation (memory) in the time series of closed and open
states fluctuations as shown in \cite{mws1}. The autocorrelation
function $k(s,t)$ of the signal $\{X_t\}_{t=1}^{T}$ is defined as
\[ k(s,t)=\frac{\left<\left(X_s-\mu_s\right)\cdot
\left(X_{s+t}-\mu_{s+t}\right)\right>}{\sigma_s\cdot\sigma_{s+t}},
\] where $\sigma_s$ is a standard deviation and $\mu_s$ is the
mean value of the sample at the moment $s$ \cite{mws1,feller}. The
stationarity of the ion current signal, which implies the
stationary character of the examined here closed and open states
fluctuations, has been discussed in \cite{ful}. The time series
has been divided into smaller sections whose similar statistical
characteristics as well as lack of any trend have been found. The
autocorrelation function of ion current signal was studied before
\cite{mws1}. Three regions of power law scaling have been
determined with two crossovers: at 1 ms and 40 ms
(Fig.~\ref{fig4}). A power law behavior has also been found in the
autocorrelation function of closed and open states fluctuations,
presented in Fig.~\ref{fig5}. For both time series the correlation
drops significantly already at $t=0.2$ ms (to the value of app.
0.3). At $t=0.5$ ms the slow power law decay begins.

Autocorrelation function of closed states fluctuations scales as

\[ k_{c} (t)\propto\left\{ \begin{array}{ccl} t^{-1.62\pm 0.40} & {\rm for }
&
t<0.5 {\rm ms}\\ t^{-0.21\pm 0.05} & {\rm for } & t>0.5{\rm ms},\\
\end{array}
\right. \]

For open states the function $k(t)$ has the following form

\[ k_{o} (t)\propto\left\{ \begin{array}{ccl} t^{-1.84\pm 0.14} & {\rm for }
&
t<0.5 {\rm ms}\\ t^{-0.37\pm 0.05} & {\rm for } & t>0.5{\rm ms},\\
\end{array}
\right. \]

Note, that the scaling exponent for $t>0.5$ ms, therefore
describing the middle- and long-range correlation, for closed
states fluctuations is very close to the value of the scaling
exponent found in the autocorrelation function of ion current
signal in the time range [1, 40] ms. The scaling exponent is in
this region equal to $0.14\pm0.02$ (see Fig.~\ref{fig4})
\cite{mws1}. This finding confirms the DFA results showing that
{\it the closed state dynamics dominates ion channel behavior} and
{\it the long range memory of the channel action is mainly due to
correlation within and between closed states}.

Having found the scaling exponents of DFA and AFA, $\alpha$ and
$\gamma$, respectively, we checked whether they are related with
each other by the known formula $\gamma =2(1-\alpha)$ \cite{bass}.
We would like to note, that those exponents can be compared with
each other only in the time region in which both data analysis
methods have some overlap, i.e. in the low time lag scaling region
($1<n<50$ ms, from Fig.3) for DFA, and the high time correlation
region ($0.5<t<40$ ms , from Fig.5) for AFA. Nevertheless, from
the definition of the $\alpha$ and $\gamma$ one should notice that
the equation relating them holds true only in the limits either
$n$ or $t$ going to 1 ms. The $\alpha$ exponents in the
overlapping region are 0.69 and 0.65, for the closed and open
states, respectively (Fig. 3). For comparison in Fig. 5, the
scaling $t^{-0.65}$ ({\bf this is just for suggesting a slope}) is
shown to indicate the expected validity of the theoretical
relationship. It appears that while $\gamma$ equals 1.62  and 1.84
for the closed and open states, respectively, from the best fit
(Pearson correlation coefficient $R^{2}=0.9$), taking into account
the 5 lowest experimental points, the value -0.65 also seems very
reliable in both cases: near the crossover, there is a remarkable
agreement for the open state case with such a value, and some
reasonable one, for the closed state case.

\section{Concluding Remarks}

The main objective of the study was to investigate correlation
properties of ion channel fluctuations within two extreme
conducting states: closed and open, eliminating cross correlation.
Previous studies showed the existence of correlations between
successive durations of closed and open states \cite{siwy}. The
procedure of determining the dwell-time series was preceded by
reconstructing the ion current time series into a dichotomous
signal; therefore the individual state fluctuations were not taken
into account. As the patch clamp technique, which enables to
determine the time course of current states, gives some information
about the conformational states of the channel, we wanted to
examine the character of the dynamics of channel conformational
substates. The closed and open states fluctuations were extracted
from ion current time series using the procedure discussed in
\cite{hu}. The approach of treating the closed (open) state as one
entity enabled us look at the "intrastate" correlation features,
occurring within one, uninterrupted state of the channel, as well
as "interstate" correlation structure, present in the state with
activity gaps. The resulting time series were subjected to DFA and
AFA methods. Both techniques show that the correlation between
sublevels of the closed states is stronger than between substates
of the open channel. Long-range correlations in the fluctuations
within the closed states are characterized by a DFA exponent close
to 1, and by a slow decay of the autocorrelation function (with a
power law exponent $0.21\pm0.05$). Both examined time series reveal
very similar scaling properties at small and medium time lag
scales: in both cases for $n<50$ ms $\alpha$ is equal to app. 0.7.

The important problem to study is the time region at which the two
methods overlap i.e. their exponents are not independent --
related by the known linear formula \cite{bass}. We have found
that for the examined data, the scaling exponents of DFA and AFA
fulfill this relation in the time scale around 1 ms. Correlation
properties for shorter lags are given by AFA, while the large time
scale is covered only by DFA. In the examined system, the two
presented methods are therefore complementary to each other.

The above results are consistent with previous analyses showing
that the long-range correlations of the BK channel functioning are
determined by the closed state correlation properties \cite{siwy}.
In this previous paper we focused on the correlation between
successive durations of closed and open states, as well as
correlation existing between subsequent dwell-times, therefore,
neighouring closed and open (Fig. 7 and 8 in \cite{siwy}). Here we
looked in detail into the structure of each state. An interesting
observation was made by comparing the results of those two
approaches. Surprisingly, the DFA analysis showed the existence of
a similar threshold at app. 50 ms, regardless whether the
sucessive durations of the closed state, or the successive actual
values of the current within the state are examined. Moreover,
this threshold was also revealed by the DFA scaling of subsequent
dwell-times durations. These findings suggest that there is a
similar mechanism, regulating intra-state behaviour of the channel
as well as duration of the dwell-times. Assuming that the observed
memory effect is mainly due to the applied field, it can be
concluded that external voltage influences the channel action on
an amazingly broad time scale, starting from the short-range
memory between the states distant by 0.1 ms, up to tens of ms
range.

An interesting problem is to consider how the macroscopic electric
field can interact with the many degrees of freedom of the channel
protein, resulting in a correlated character of the protein
dynamics. This question is related to the main function of the
voltage-gated family of channels, i.e. stabilization of the
transmembrane voltage, which is certainly influenced by the
channel capability to close and to open. This question was
examined in the multi-channel system consisting of voltage
dependent {\it Shaker} channels, expressed in the membrane of a
{\it Xenopus} oocyte \cite{salman1,salman2,braun}. The system was
studied by observing voltage fluctuations for a various number of
channels. For a low number of channels (the smallest number
reported was 19) the observed fluctuations revealed a wide range
of time scales, which pointed to the existence of multiple closed
and open states. We focused here on the dynamics of a single
voltage-gated channel, (studied on the basis of ion current
fluctuations), and showed the {\it self-similarity} of both closed
and open states independently of each other. Our results point to
the necessity of performing a thorough analysis of the dynamics of
a single channel, before creating models of the possible
connection between microscopic dynamics of the channel protein and
macroscopic biological phenomena such as transmembrane potential.
Indeed, we show that the most important outcome of this study is
to show the existence of dynamical correlation already in a
single-channel, with two conducting states.

\subsection{Acknowledgments}

We are grateful to Professor P.N.R. Usherwood and Dr I. Mellor
from the University of Nottingham for providing us with the
experimental data of ion current through a large conductance
locust potassium channel. The authors are grateful to Prof. Karina
Weron, Dr Szymon Mercik and Hilmar Spohr for discussions and
careful reading of the manuscript.

Z.~S. greatly acknowledges financial support from the Foundation
for Polish Science; she was granted a one-year post-doc fellowship
at the Gesellschaft f\"ur Schwerionenforschung in Darmstadt,
Germany.

\newpage 

\end{document}